\documentclass[aps,tightenlines,nofootinbib,preprint,
superscriptaddress,groupedaddress,epsfig]{revtex4}
\usepackage{graphicx}
\newcommand{\be}{\begin{equation}}
\newcommand{\ee}{\end{equation}}
\newcommand{\bear}{\begin{eqnarray}}
\newcommand{\eear}{\end{eqnarray}}
\begin{document}

\title{Resummed pole mass, static potential, and  $\Upsilon$(1S) spectrum}

\author{C. S. Kim}
\email{cskim@yonsei.ac.kr}
\affiliation{Department of Physics, Yonsei University, Seoul 120-479, Korea}
\author{Taekoon Lee}
\email{tlee@phya.snu.ac.kr}
\affiliation{Department of Physics, Seoul National University,
Seoul 151-742, Korea}
\author{Guo-Li Wang}
\email{glwang@cskim.yonsei.ac.kr}
\affiliation{Department of Physics, Yonsei University, Seoul 120-479, Korea}


\begin{abstract}
Employing the heavy quark pole mass and static potential that were
obtained from resummation of the renormalon-caused large order
behavior to all orders, and matching  the resummed static
potential to the long distance Cornell potential we obtain a new
estimate of the bottom quark mass 
from $\Upsilon$(1S)  system. We point out that the pure perturbative QCD
approach to $\Upsilon$(1S) spectrum should have a sizable systematic error
of about 100 MeV, due to the absence of the long distance
confining potential, and that the incorporation of the
latter is essential for accurate determination of the ground state
energy. We also recalculate the nonperturbative Leutwyler-Voloshin
effect using a resummed octet potential matched to a confining
linear potential, and obtain a result that is substantially
smaller than the  estimates based on Coulombic
potentials. The hyperfine splitting is estimated to be $50 \pm 8$
MeV at $\alpha_s(M_z)=0.118$.
\end{abstract}



\maketitle

\section{Introduction}

The heavy quarkonium is a quantum chromodynamic (QCD) system that essentially
depends on the two fundamental parameters, the heavy quark mass
and strong coupling constant. The precise experimental
determination of the spectra of heavy quarkonium such as
bottomonium provides a good opportunity to test QCD as well as
determine the fundamental parameters.

Traditionally, the heavy quarkonium was investigated primarily within the
phenomenological potential models that proved very successful. Nevertheless,
the ambiguities  in the potential models forbade a precise connection
to QCD, and  thus an  accurate determination of the QCD parameters in this
approach was not feasible. The  recent approaches to quarkonium system is based
on effective  theories such as the potential NRQCD (pNRQCD) \cite{pnrqcd}.
In pNRQCD
Lagrangian the heavy
quark mass, which appears in its most natural form as a pole mass, and
the interquark potentials are matching coefficients, which can
be calculated, formally, in perturbative QCD (pQCD).

For the bottomonium the first energy eigenstates are expected to be
largely determined
by the interquark potentials at short distances which can be calculated
in pQCD. If the nonperturbative effects are indeed small then
these first states are  essentially perturbative, and can open a window for
testing  pQCD as well as determining the QCD parameters.
Crucial for the success of this approach is
that the nonperturbative effects be small and the  potentials at short
distances as well as the pole mass be calculable reliably within pQCD.
It is, however, well known that because of the infrared (IR) renormalons the
perturbative expansions for the pole mass as well as the static
potential, which are known to next-next leading order (NNLO),
show bad convergence even at scales where pQCD should be applicable.

Besides the apparent convergence problem the IR renormalons pose a
more fundamental problem of defining the infrared-sensitive
quantities. Let us focus first on the pole mass for the moment,
but the discussion in the following applies equally well to the
static potential. The renormalon divergence in the perturbative
expansion of the pole mass causes an intrinsic uncertainty of
${\cal O}(\Lambda_{\text QCD})$, and this results in unavoidable
ambiguity in defining the pole mass. Nevertheless, a pole mass can
be well defined   by Borel resummation. In Borel resummation the
renormalon ambiguity appears as an ambiguity in taking the
integration contour over the renormalon singularity on the Borel
plane. Despite this ambiguity one can  define a pole mass simply
by taking the principal value prescription for the Borel integral.
We call the pole mass defined this way the BR mass
\cite{lee1,lee2}. The BR mass, by definition, has exactly the same
perturbative expansion as the  pole mass in pQCD, and thus the
coefficients suffer from the same renormalon divergence, but
nonetheless it is well-defined. Obviously, the BR mass is not a
short-distance mass, and in this respect it is unique among the
various heavy quark mass definitions. The ambiguity in defining
the pole mass cancel out in physical obeservables, for instance in
heavy quarkonium  Hamiltonian the ambiguities in the pole mass and
the static potential cancel each other, leaving the Hamiltonian
free from renormalon divergence \cite{neubert,hoang,beneke}. It is
noted that we focus, throughout the paper, exclusively on the
first  IR renormalon since it is overwhelmly dominant in the
divergence of the perturbative coefficients of the pole mass as
well as the static potential.

Although the Borel resummation formally resolves the renormalon
problem, practicing it is another matter since the calculation of
the Borel integral to an accuracy better than the renormalon
ambiguity demands the Borel transform be known precisely in the domain
containing the
origin as well as the first IR renormalon singularity.
Although the behavior of the Borel transform about
the origin can be obtained by the ordinary perturbative expansions,
obtaining its behavior from the origin to the renormalon singularity
 with the known first terms of perturbation is
nontrivial, since it requires  accurate knowledge on the
renormalon-caused large order behavior. This  difficulty of Borel
resummation led in the past to the abandonment of an independent
notion of infrared sensitive quantities such as the pole
mass, and to the adoption of the short-distance quantities instead
\cite{beneke1}.

Since the difficulty of achieving a manageable definition of an
infrared sensitive quantity arises from the difficulty of
describing  the Borel
transform about the renormalon singularity, it is worthwhile to
look at this problem more carefully. As  is well known, the exact
nature of the renormalon singularity can be determined by
renormalization group (RG) equation, and the only missing
information about the renormalon singularity
is the residue that determines the overall
normalization constant of the large order behavior. Although the
residue cannot be calculated  exactly, there is a perturbation
scheme in which the residue can be expressed as a convergent
series that can be calculated using the ordinary perturbative
expansions \cite{lee97,lee99}.
The scheme, when applied to  QCD observables, yields
residues that show surprisingly  good convergence
even with the first terms of
the perturbative expansions, which indicates the residues can be
calculated fairly well from the low order perturbative
calculations. It was observed recently that this
method works particularly well for the  pole mass
and the static potential because of the overwhelming dominance
of the first IR renormalon in these quantities
\cite{pineda01,pineda02,leepotential}.

With the residue known approximately, one can obtain a good
description of the Borel transform in the region that contains the
origin and the renormalon singularity through an interpolation of
the perturbative expansions about the origin and the functional behavior about
the renornalom singularity. A particular prescription of the
interpolation called {\it bilocal expansion} was proposed in
\cite{lee3}, which
 incorporates the correct nature of the
renormalon singularity and gives, by definition, exactly the same
perturbative expansion about the origin as the true Borel
transform. The Borel resummation employing this interpolating Borel
transform
then sums the renormalon-caused large order behavior to
all orders, giving an all-order definition for the infrared sensitive
quantity.
The pole mass and
static  potential computed in this scheme show
remarkable convergence under the bilocal expansion, and in the case
of the latter an excellent agreement was obtained with the lattice potential
\cite{lee3}.
With the Borel summation technique we thus have a
well-defined pole mass that has exactly the same perturbative
expansion as the  pole mass in pQCD, but nonetheless converges
rapidly under the bilocal expansion.

In this paper we apply the Borel resummed pole mass and  the static
potential to $\Upsilon$(1S) system to determine the bottom quark mass and
the hyperfine splitting. Our approach is unique in that it is not based on
short-distance quantities.
An useful feature of employing the all-order quantities is that a natural scale
separation occurs in a multi-scale system such as the quarkonium.
In the schemes that employ short-distance quantities  the large and small
scales mix invariably in the perturbative calculations, causing large
perturbative coefficients that render the calculations less reliable.
Since in our scheme the pole mass and
the potential can be resummed at scales that are optimal for each quantities,
for example, the heavy quark mass and the inverse of
the interquark distance, respectively, the Hamiltonian in our
approach does not suffer from the scale mixing problem. This is a feature
that differentiates our scheme from those based on
short-distance quantities.
There are other advantages, for instance, having a well-defined, stand-alone
potential makes it easy to merge
the resummed potential at short distance to the phenomenological
long distance confining potential.

The paper is organized as follows. In Sec. II, we define the
Hamiltonian in BR scheme that is obtained, at short and
intermediate distances, rigorously from pQCD plus resummation and
at intermediate and long distance by matching the long distance
Cornell potential to the BR potential. In Sec. III, we show that
the incorporation of the long distance confining potential is
important for $\Upsilon$(1S) and the computations based on pure
pQCD should contain sizable systematic error. In Sec. IV, we
recalculate the Leutwyler-Voloshin type nonperturbative effects
employing the resummed singlet and octet potentials and show that
they are sufficiently small enough to be negligible for the 1S
state. In Sec. V we give our estimate of the $\overline{\rm MS}$
bottom quark mass and hyperfine splitting, and in Sec. VI give a
summary.

\section{Hamiltonian}

The pNRQCD Lagrangian for heavy quarkonium contains singlet as
well as octet fields and ultrasoft gluons which represent the
dynamical degrees of freedom at scales below the typical heavy-quark
momentum. Ignoring for the moment the interactions between the
singlet and octet fields, the pNRQCD Hamiltonian for singlet
quarkonium in on-shell scheme is given by \cite{pnrqcd-h}
\bear H&=& H_0
+H',\label{eq1} \eear where \bear H_0&=& 2 \,m_{pole}
-\frac{\nabla^2}{m_{pole}} + V^{(0)}(r)\,, \qquad H'\equiv
\sum_{n=1} \frac{V^{(n)}}{m_{pole}^n} \,. \eear
For the 1S state
$H'$, in the leading order, reads \cite{subleading,sumino}
\be H'=V_{NA}+
V_{hf} + V_{SI} \label{eq2}\,, \ee
with the  nonabelian and hyperfine
terms given by \bear V_{NA} = - \frac{C_A C_F
\alpha_s(\mu)^2}{2m_br^2}\,, \qquad V_{hf} = \frac{8\pi C_F
\alpha_s(\mu)}{3m_b^2} \delta^3(\vec{r})\,, \label{eq4} \eear
 and   the spin--independent operator by
\begin{equation} V_{SI} =  - \frac{\vec{p}\,^4}{4m_b^3} \, +
\frac{\pi C_F \alpha_s(\mu)}{m_b^2} \, \delta^3(\vec{r}) - \frac{C_F
\alpha_s(\mu)}{m_b^2r}\vec{p}\,^2 \,. \label{eq5}
\end{equation}

We shall now focus on the zeroth order Hamiltonian $H_0$.
As mentioned the pole mass $m_{pole}$ and singlet
potential $V^{(0)}$ in $H_0$
are ill-defined in pQCD because of the
renormalon divergences in the perturbative matching
relations \cite{shifman,bb} 
\bear
m_{pole}&=&m_{\overline{\rm MS}}(1+\sum_{n=0}^\infty
p_n(m_{\overline{\rm MS}},\mu)
\alpha_s(\mu)^{n+1} )\,, \nonumber \\
V^{(0)}(r)&=& -\frac{1}{r}\sum_{n=0}^{\infty} V^{(0)}_n(r,\mu)
\alpha_s(\mu)^{n+1}\,, \label{eq7} \eear
where $\alpha_s(\mu)$ denotes
the strong coupling constant.

The Hamiltonian in BR scheme can be obtained by substituting
$m_{pole},V^{(0)}(r)$ with the Borel resummed quantities $m_{\rm BR},
V^{(0)}_{\rm BR}(r)$ which are defined by \cite{lee1,lee2,leepotential} \bear
m_{\rm BR}(r)&=& m_{\overline{\rm MS}}\left(1+
Re\left[\frac{1}{\beta_0}\int_{0+i\epsilon}^{\infty+\epsilon}
e^{-b/\beta_0\alpha_s(m_{\overline{\rm MS}})} \tilde
m(b,m_{\overline{\rm MS}}) d\,b
\right]\right)\,,\nonumber \\
V^{(0)}_{\rm BR}(r)&=& Re\left[\frac{1}{\beta_0}
\int_{0+i\epsilon}^{\infty+\epsilon}
e^{-b/\beta_0\alpha_s(1/r)} \tilde V^{(0)}(b,1/r) d\,b \right]\,.
\label{eq8}
\eear
The Borel transforms in these equations have
perturbative expansions about the origin,
\bear \tilde m(b,m_{\overline{\rm MS}})&=&\sum_{n=0}
\frac{p_n(m_{\overline{\rm MS}},m_{\overline{\rm MS}})}{n!}
\left(\frac{b}{\beta_0}\right)^n\,,
\nonumber \\
\tilde V^{(0)}(b,1/r)&=&\sum_{n=0} \frac{V^{(0)}_n(r,1/r)}{n!}
\left(\frac{b}{\beta_0}\right)^n\,,\label{eq9} \eear
where $\beta_0$ denotes the one loop coefficient of the QCD $\beta$ function,
and about the renormalon singularity:
\bear
\tilde m(b,m_{\overline{\rm MS}})& =& \frac{C_m}{(1-2b)^{1+\nu}}\left[
1+c_1 (1-2b)+c_2(1-2b)^2+{\ldots} \right] + \text{ (analytic part)}\,,\nonumber
\\
\tilde V^{(0)}(b,1/r) &=& \frac{C_v}{(1-2b)^{1+\nu}}\left[ 1+c_1
(1-2b)+c_2(1-2b)^2+{\ldots}\right] + \text{ (analytic part)}\,,
\label{e99}\eear
where $\nu$ and $c_1,c_2$ are exactly known
constants. The renormalon residues $C_m, C_v$ can be computed
perturbatively following Refs. \cite{lee97,lee99}. Combining the
expansions Eq. (\ref{eq9}) and Eq. (\ref{e99}) in bilocal
expansions, a good description of the Borel transforms can be
obtained from the known NNLO perturbations in Eq. (\ref{eq7}). For
details we refer the readers to \cite{lee1,lee2,leepotential}.

Note that in the Borel summations (\ref{eq8}) we have chosen
separate scales, $\mu=m_{\overline{\rm MS}}$ for the pole mass and
$\mu=1/r$ for the potential, which are, respectively, the optimal
scales for the perturbative expansions (\ref{eq7}).
Since the renormalon ambiguities must be RG invariant we can choose the
renormalization
scales independently, without spoiling the cancellation of
the renormalon ambiguities between the pole mass and the static potential.
This is the crucial feature of our Borel
resummation approach.

\begin{figure}
\begin{picture}(400,100)(0,0)
\put(0,0){\includegraphics{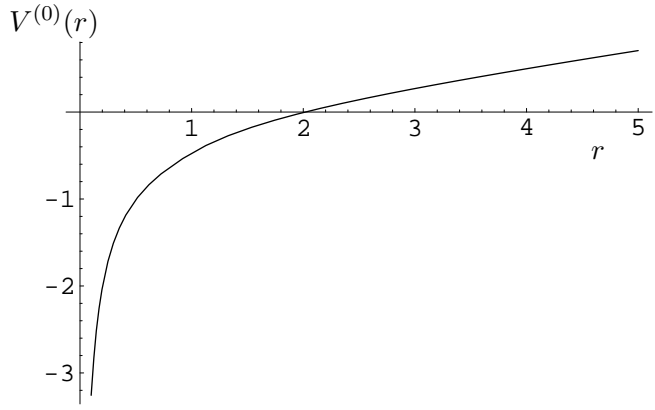}}
\put(300,100){\makebox(2,2)[l]{$r$}}
\put(80,150){\makebox(2,2)[l]{$V^{(0)}(r)$}}
\end{picture}
\caption{The improved potential $V^{(0)}(r)$ obtained by matching
the BR potential and Cornell potential.
}
\end{figure}

Although the Borel resummed potential is expected to give a good
description at short distance it will fail at long distance. To
obtain the potential at all distances we follow the old  idea of interpolating
the pQCD potential to the confining linear potential. There are many
implementations of this idea, from the original Richardson  potential to
the recent one based on renormalon cancellation \cite{sumino}.
Our implementation is as follows. We combine
the BR potential valid at short distances with the Cornell
potential at long distance by matching them at an intermediate
distance $r_0$ by \bear V^{(0)}_{\rm BR}(r_0)=V_{\rm Cornell}(r_0)\,,
\quad \frac{d}{dr} V^{(0)}_{\rm BR}(r_0)=\frac{d}{d r}
V_{\rm Cornell}(r_0) \,,\label{eq10} \eear where \be V_{\rm Cornell}(r)= a
+\frac{b}{r} + c r\,. \label{eq11}\ee
Our choice of the Cornell
potential as the long distance potential is based not only on its
phenomenological success but its capability to fit the lattice
data well \cite{sommer}. We take the linear slope as  $c=0.18\,\,
{\rm GeV}^2$, which is motivated by the lattice calculation \cite{bali}.
Note that the matching equations completely determine the two free
parameters in the Cornell potential. Thus our hybrid singlet potential
reads \bear V^{(0)}=\left\{ \begin{array}{ll}
                 V^{(0)}_{\rm BR}(r)\,, &\quad  r \leq r_0\,, \\
         V_{\rm Cornell}(r)\,,  &\quad  r \geq r_0\,.
         \end{array}
         \right.\label{eq12}
\eear
The potential obtained this way is continuous and
smooth (see, Fig. 1) at the matching point  $r_0$.

The Borel resummed potential in pure QCD ($n_f=0$) agrees well with
lattice calculations to distances as large as $r\approx (700 {\rm MeV})^{-1}$
\cite{leepotential}.
Assuming that this behavior does not change drastically as the light quarks are
introduced, we shall take  the matching scale  to be $r_0^{-1}=1$ GeV.
It will be shown that the dependence of the $\Upsilon$(1S)
binding energy on the matching
scale is very small.

As we shall see in the next section  the incorporation of the long
distance confining potential is important even for such a
spatially small system as $\Upsilon$(1S), whose wave function turns out to
have a sizable support where the difference between the confining
potential and the perturbative potential is substantial. This
implies that $\Upsilon$(1S) is not completely perturbative even
without the ultrasoft effects that will be discussed in Sec. \ref{LV}.

\begin{figure}
 \includegraphics[angle=0 , width=10cm
 ]{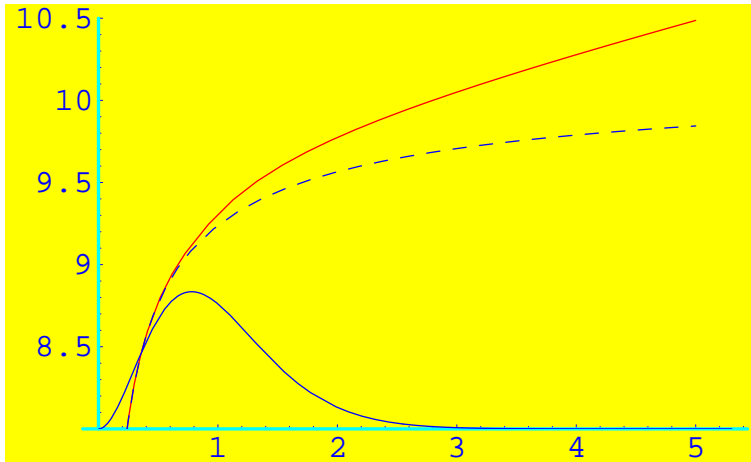}
\caption{The total energy in BR method (solid line), pQCD method
(dashed line) and zeroth order wave function $r^2\Psi^{(0)}(r)^2$ shifted by
$8$ for displaying purpose. }
\end{figure}

\section{Systematic error in pure perturbative approach}

Recently there have been higher order calculations of the
$\Upsilon$(1S) spectrum within pure pQCD framework (see for
example, \cite{subleading,bvy,penin,meta}). In this approach the
leading order binding energy can be obtained by solving the
Schroedinger equation of the Hamiltonian, obtained by the
perturbative expansion of $2m_{pole}+ V^{(0)}(r)$ in $H_0$ using
the power expansions (\ref{eq7}) at a fixed renormalization scale
$\mu$. Because of the renormalon cancellation between $2 m_{pole}$
and $V^{(0)}$ the perturbative expansion of the spectrum is free
from renormalon divergence (note that, throughout the paper,
divergences due to the subleading renormalons are ignored).
However, this approach can have a significant systematic error
that cannot be seen from the convergence of the perturbative
expansion or the renormalization scale dependence of the computed
spectrum, since $2 m_{pole}+ V^{(0)}(r)$, although converges
perturbatively, cannot produce the confining potential however
high order it has been calculated.

\begin{table*}[hbt]
\caption{\footnotesize The zeroth order total energy $E^{(0)}$
by pure perturbative expansion method with differing set of
scale $\mu$, the unit is in GeV.}
\begin{ruledtabular}
\begin{tabular}{lccccc}
$\mu$&1&2&3&4&5 \\ \hline
$E^{(0)}$&9.383&9.486&9.411&9.413&9.390\\
\end{tabular}
\end{ruledtabular}
\end{table*}

To test whether the incorporation of the confining potential is
important for $\Upsilon$(1S) state, we calculate the ground state energy
of $H_0$ in BR as well as the perturbative schemes and
compare them. We first plot $2 m_{pole}+ V^{(0)}(r)$ in Fig. 2.  in
both BR scheme and in pQCD using NNLO calculations of the
expansions (\ref{eq7}).
We put $m_{\overline{\rm MS}}=4.2$ GeV and $\alpha_s(M_Z)=0.118$
for both BR and pQCD schemes.
In pQCD case the pole mass in the kinetic term were substituted  by the
NNLO pole mass in Eq. (\ref{eq7}).
The plot shows that the difference between these two schemes becomes
substantial at $r \geq 1 {\rm GeV}^{-1}$, where the wave function for
$\Upsilon(1S)$ has significant support, and because of this the
computed
spectrum is sensitive on the employed method. We list
the  spectrum in Table I computed  by the pQCD method. Since the
BR result is $E^{(0)}=9.574$ GeV, we find the energy difference,
which depends on
the renormalization scale $\mu$, is at least $100$ MeV.
This implies that the achievable accuracy in an approach based on pure pQCD
with a fixed renormalization scale is at best 100 MeV, and that an
incorporation of the confining potential at long distance
is essential for precision calculation of $\Upsilon$(1S) spectrum.

\section{\label{LV}
Recalculation of the Leutwyler-Voloshin effects}

The binding energy of the color singlet quarkonium is also
affected by the emission and absorption of the ultrasoft gluons,
which occur in pNRQCD through the singlet-octet-gluon vertex. This
effect is nonperturbative and difficult to calculate.
However, when the typical time scale of the quarkonium, $T\sim 1/mv^2$,
 is much smaller than the vacuum
fluctuation scale, $1/\Lambda_{QCD}$, the quarkonium would `see'
the ultrasoft gluon as a constant random  background field.  The
ultrasoft effect then becomes the Stark effect which in the case of
quarkonium was first investigated by Voloshin \cite{voloshin} and
Leutwyler \cite{leutwyler}.

Whether this Stark effect approximation is applicable to $\Upsilon$(1S)
is debatable, but nonetheless we shall assume in the following
it is the case.  As argued
in Ref. \cite{gromes}, even if this assumption fails, it is however likely
that the Stark effect approximation will give an upper bound on
the the ultrasoft effect, since
the local condensate approximation  gives an upper bound on the
the two-point correlation function of the gluon field strength tensors.

According to Leutwyler and Voloshin the Stark effect can be
large on $\Upsilon$(1S) state and  this  is known to give a
large uncertainty  to the spectrum obtained in pQCD approach.
The  calculations by Leutwyler and Voloshin employed
Coulombic potentials for the singlet as well as octet channels.
Obviously, these potentials must not be good representations of the
true potentials, and thus the validity of the  calculations
based on Coulombic potentials is questionable.

Our scheme of matching the Borel resummed potential at short distance with
the confining potential at long distance is well-suited for the re-examination
of this problem.
In this section we
recalculate the Stark effect using the  hybrid singlet and octet  potentials
that were obtained by interpolations of the Borel resummed potential at short
distance potentials and  the confining potentials at long
distance.

Surprisingly, our result, which will be given shortly, shows that
the Stark effect in $\Upsilon$(1S)
state is very small. This is fortunate since it removes the large
uncertainty  coming from the ultrasoft nonperturbative effect.

The nonperturbative correction to the ground state energy is given
by
\bear \delta E_{NP}= \frac{\pi}{18}\langle\alpha_s
G_{\mu\nu}^2\rangle \sum_{i=1}^3 \langle\Psi_0|r^i \frac{1}{\hat
H_0-E_0}r^i|\Psi_0\rangle\,, \label{eq13} \eear
where $\Psi_0, E_0$ are,
respectively, the wave function and energy of the ground state of
$H_0$, and $\hat H_0$ denotes the Hamiltonian for the octet
channel: \bear \hat H_0= 2 m_{\rm BR} -\frac{\nabla^2}{m_{\rm BR}} +
V_O(r)\,, \label{eq14} \eear
where $ V_O(r)$ denotes the static octet
potential.

With a recent preliminary calculation of the NNLO term \cite{nnlo-octet}
the octet potential is known perturbatively to NNLO.
At short distance it is
repulsive, but at long distance the octet potential
is expected to have a confining
linear potential. A hybrid octet potential that has good short distance
behavior and a confining potential at long distance can be
constructed in a similar fashion for the singlet potential. We
first obtain the Borel resummed short distance potential employing
the NNLO bilocal expansion, and then match it at the radius
$r_0=1 {\rm GeV}^{-1}$ to the fitted  potential $V(r)=d +\sqrt{3.543-0.289\,
r+0.077\,r^2}$, which is obtained by fitting the lattice data for
the octet potential \cite{kuti}. The matching condition is that
the potential be continuous at $r=r_0$, and this determines the
constant $d$. Note that the asymptotic linear slope of the fitted
potential $\sqrt{0.077}\approx0.277\,{\rm GeV}^2$ is roughly consistent with the
bag model calculation, according to which the slope of the octet
potential is given by $\sqrt{7/4}$ times that of the singlet
potential \cite{bag}. The octet potential obtained this way is
plotted in Fig. 3. Now the nonperturbative correction
Eq. (\ref{eq13}) can be easily evaluated by numerically solving the
equation \bear (\hat H_0-E_0)|\chi^i_0>=
r^i|\Psi_0>\,.\label{eq15} \eear

\begin{figure}
 \includegraphics[angle=90 , width=10cm]{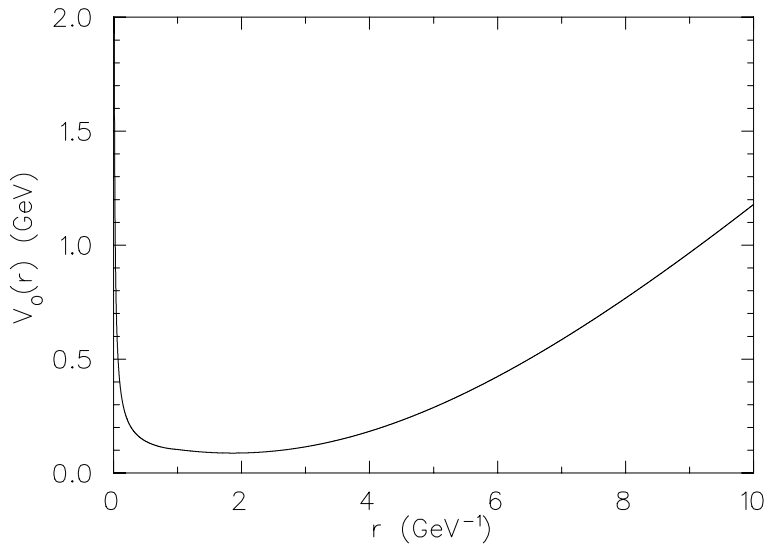}
 \caption{The octet potential, which is obtained by the
 Borel resummation at $r<1$ GeV$^{-1}$ and by fitting the lattice data
 at $r>1$ GeV$^{-1}$. }
\end{figure}

\section{Results}

To obtain the binding energy we first solve the
Schr\"odinger equation for $H_0$ numerically to obtain the zeroth order
binding energy $E^{(0)}$ and the wave function $\Psi^{(0)}$, and
treat the interactions
$V_{NP}, V_{hf}$ and $V_{SI}$  as perturbations.

\begin{figure}
\begin{picture}(400,100)(0,0)
\put(0,0){\includegraphics{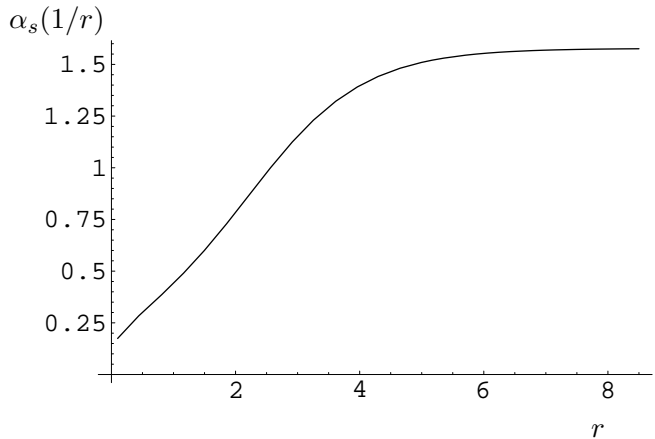}}
\put(300,-5){\makebox(2,2)[l]{$r$}}
\put(80,150){\makebox(2,2)[l]{$\alpha_s(1/r)$}}
\end{picture}
\caption{Regularized $\alpha_s(1/r)$. In infrared region
$\alpha_s(1/r)$ tends to a fixed value. }
\label{alpha}
\end{figure}

Before performing the calculation, we have to deal with two
problems caused by our choice of the inverse of the interquark
distance as the renormalization scale for the potential.  The
hyperfine and spin-independent operators in Eqs.
(\ref{eq4},~\ref{eq5}) contain the delta function
$\delta^3(\vec{r})$, which will result in vanishing couplings
$\alpha_s[1/({r \rightarrow 0})]$ in the numerator. In this case
the inverse of the interquark distance as renormalization scale is
clearly inappropriate. We solve this problem by noting that the
renormalization scale  for $V^{(0)}(r)\,,\,V^{(1)}(r)$ can be
independently chosen from the rest terms in $H'$ since these two
terms should be RG invariant, independently from the rest terms of
$H'$. We therefore choose the renormalization scale $\mu=1/r$ for
$V^{(0)}$ and $V_{NA}$ and a fixed value of $\mu$ for the
hyperfine and spin--independent terms.
 The second problem is caused by
the value of $\alpha_s(1/r)$ at the opposite condition, when $r$ is large.
The coupling $\alpha_s(1/r)$ obtained by the perturbative $\beta$ function
is not reliable when $r$ is large. To avoid this
difficulty, one can use a simple regulation \cite{peskin}
\begin{equation}r \Lambda_{QCD} \to a \times{\rm tanh}\left(\frac{r
\Lambda_{QCD} }{a}\right)\,, \label{eq16}\end{equation}
in the running coupling $\alpha_s(1/r)\equiv f(r\Lambda_{QCD})$, which is
obtained using the four-loop $\beta$ function in $\overline{\rm MS}$ scheme.
The free parameter $a$ should not be chosen too small a value so that
the regularized coupling does not deviate too much from the perturbative
coupling
at short distances, nor too large a value to avoid an unphysical dip in the
vicinity of the Landau pole.
In our calculation,
we set $a=0.55$. We shall see that the computed $\Upsilon$(1S) spectrum
has only a mild dependence  on this value.  With this choice
 the regularized coupling changes little from the perturbative coupling
 when $r$ is smaller than $r=1$ GeV$^{-1}$, and at large $r$
 the regulation suppresses the growth of the perturbative coupling and
 approaches to an infrared fixed value $\alpha_s(0)\approx 1.5$ (see Fig.
\ref{alpha}).
 This infrared behavior of the regularized coupling is interesting, considering
 that the effective running couplings in various dispersive approaches display
 freezing behavior in the infrared limit \cite{infrafixed}.
 Incidently,   our
 $\alpha_s(0)\approx 1.5$ is  very close to the predicted infrared fixed value
$1/\beta_0=1.507\,\, (\text{at} \,\, n_f\!=\!4)$ of the effective
coupling in ``analytic perturbation theory'' \cite{apt}.

Setting the input values of the strong coupling constant
$\alpha_s(M_z)=0.118$, we solve the non-relativistic Schr\"odinger
equation numerically with the hybrid potential, obtain the zeroth
order binding energy $ E^{(0)}$ and wave function $\Psi^{(0)}$. We
treat the interactions $V_{NA}$, $V_{hf}$ and $V_{SI}$ as
perturbations and compute the perturbative corrections by
calculating the expectation value
\begin{equation} \delta E = \langle {\Psi^{(0)}}\mid
( V_{NA} + V_{hf} + V_{SI})
 \mid{\Psi^{(0)}}\rangle .\label{eq177}
\end{equation}
The contribution from each term  is  given in Table \ref{dele}. The
values for the hyperfine and spin-independent contributions are
with the renormalization scale $\mu=3$ GeV.

The  nonperturbative Leutwyler-Voloshin effect defined in
Eq. (\ref{eq13}) is computed by solving the equation (\ref{eq15}).
Taking the gluon condensate $\langle\alpha_s
G_{\mu\nu}^2\rangle=0.02\pm 0.02$ we find $\delta E_{NP}=5.8 \pm
5.8$ MeV, which is almost an order of magnitude smaller than estimates
based on Coulombic potentials, which range $30--100$ MeV.
We note that the smallness of this
nonperturbative effect is very robust, depending little on the
shape of the octet potential above the matching distance $r_0$. We
also report, without details, that Leutwyler-Voloshin effect on
the hyperfine splitting is very small, at most $1.6$ MeV. This
result indicates that the ultrasoft nonperturbative effect, which
is regarded a major difficulty in precision computation of
quarkonium spectrum, is small in $\Upsilon$(1S) state.

Adding all these contributions as well as the charm mass effect,
which shifts the binding energy about $-20 \pm 15$ MeV
\cite{lee1}, to fit the measured mass of $\Upsilon$(1S), we obtain
the bottom quark BR  mass, $m_{\rm BR}=4.901$ GeV, and the
corresponding ${\overline {\rm MS}}$ mass $m_{\overline {\rm MS}}=4.211$ GeV.

It is noteworthy that the hyperfine and spin--independent
contributions cancel largely. The sum of these two contributions
is $-20.5 + 30.4 \alpha_s (\mu)$, in MeV, which shows that it has
only a small renormalization scale dependence. Thus, our prediction of the
$b$-quark mass depends little on the renormalization scale chosen for the
hyperfine and spin--independent terms.

Our prediction of the hyperfine splitting can be found in Table
\ref{table3} and Table \ref{table4}. This result includes the leading as
well as the subleading perturbative corrections for the hyperfine
splitting \cite{buchmueller}. At the leading order the hyperfine splitting
is quite sensitive on the value of the renormalization scale, but when
the subleading correction is included it has a very small renormalization scale
dependence
for $\mu \geq 2 $ GeV. Taking $\mu=3$ GeV and $\alpha_s(M_z)=0.118$, and
the subleading correction as the theoretical uncertainty, we
estimate the hyperfine splitting to be $50 \pm 8$ MeV.

\begin{figure}
 \includegraphics[angle=0 , width=8cm
 ]{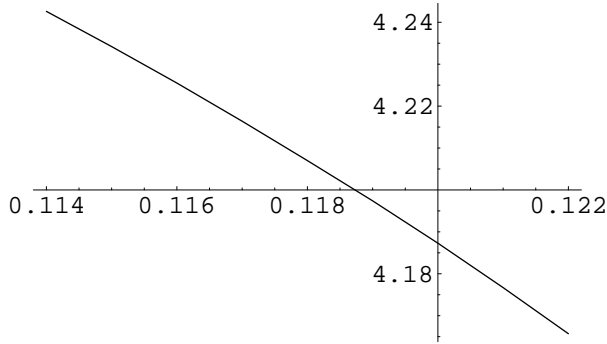}
\caption{\label{tbh} 
The extracted $b$-quark ${\overline {\rm MS}}$ mass vs  $\alpha_s(M_z)$.}
\label{massvsalphas}
\end{figure}

In the following we discuss on the uncertainties on the extracted
$b$-quark mass in our scheme. (i) Uncertainty of $\alpha_s(M_z)$:
With different input values of $\alpha_s(M_z)$, the extracted  BR
mass as well as ${\overline {\rm MS}}$ mass will change. Fig.
\ref{massvsalphas} shows the estimates of $b$-quark ${\overline
{\rm MS}}$ mass against $\alpha_s(M_z)$. We can see from the
figure, if the value of $\alpha_s(M_z)$ vary from $0.117$ to
$0.120$, the  ${\overline {{\rm MS}}}$ mass will vary from $4.220$
GeV to $4.192$ GeV. (ii) Uncertainty of $r_0$: Varying the
matching point $r_0$, at which the phenomenological potential is
stitched to the  perturbative BR potential, will cause some
uncertainty. With about $10\%$ variation of $r_0$, the extracted
$b$ quark $\overline{\rm MS}$ mass varies only within $3$ MeV.
This uncertainty can be completely ignored. (iii) Uncertainty of
$\lambda$: In our calculation, we chose the slope of the linear
potential $\lambda=0.18$ GeV$^2$. We shall assign $10\%$ variation
on $\lambda$, which causes the $\overline{\rm MS}$ $b$-quark mass
vary $4.211 \pm 0.001$. Again the error is small enough to be
ignored. The $\Upsilon$(1S) state is spatially small enough that
its spectrum is not so sensitive on the shape of the confining
potential. (iv) Uncertainty of the $\alpha_s$ regularization
constant $a$: The regularization equation (\ref{eq16}) of
$\alpha_s$ involves the constant $a$, on which the infrared fixed
value of $\alpha_s(1/r)$ depends strongly. For instance when we
vary the value of $a$  from $0.35$ to $0.75$, about $36\%$
variation from the central value $a=0.55$, the infrared fixed
value of $\alpha_s(1/r)$ vary from 0.58 to 28.6, respectively.
However, the variation of the extracted $\overline{\rm MS}$
$b$-quark mass is negligibly small, only from $4.205$ to $4.217$.
This is similar to the small dependence on the slope $\lambda$.
Since only the long distance behavior of the regularized coupling
is sensitive on $a$ the spectrum of the $\Upsilon$(1S) has only a
small dependence  on the value of $a$.

\begin{table*}[hbt]
\setlength{\tabcolsep}{0.5cm} \caption{\small Contributions
to $1S$ state in unit of MeV. }
\begin{tabular*}{\textwidth}{@{}c@{\extracolsep{\fill}}ccccc}
 \hline \hline
 {\phantom{\Large{l}}}\raisebox{+.2cm}{\phantom{\Large{j}}}
  $E^{(0)}$&$V_{SI}$&$V_{NA}$ &$V_{hf}$&$\delta E^{(0)}_{NP}$ \\
 \hline
 {\phantom{\Large{l}}}\raisebox{+.2cm}{\phantom{\Large{j}}}
   $9575$&$-$66 &$-$113 &58&6 \\\hline
 \hline
\end{tabular*}
\label{dele}
\end{table*}
\section{Summary}

We investigated the $\Upsilon$(1S) system using a hybrid Hamiltonian
for heavy quarkonium, which was built by matching the Cornell potential
at long distances to the Borel resummed pQCD potential at short
distances. This matching fixes completely the free parameters of the
Cornell potential except for the slope of the linear term.
The hybrid Hamiltonian employs an infrared sensitive,  all-order
resummed pole mass and static potential, and is free from the inevitable
scale mixing problem present in approaches based on short-distance quantities.

The incorporation of the long distance confining potential was shown to be
essential for precision calculation of $\Upsilon$(1S) system. The
$\Upsilon$(1S) is  not a completely perturbative
system, even without the ultrasoft contributions. We note that
 pure pQCD calculations of $\Upsilon$(1S) spectrum at a fixed renormalization
 scale should contain a sizable systematic error of about 100 MeV.

The nonperturbative ultrasoft effects were reconsidered in the
modeling of Leutwyler and Voloshin, employing the hybrid singlet
and octet potentials. We find these effects are very small,
contributing at most 6 MeV to the $\Upsilon$(1S) spectrum, and
negligible contribution to the hyperfine splitting which was found
to be $50 \pm 8$ MeV at $\alpha_s(Mz)=0.118$.

\begin{table*}[hbt]
\caption{\footnotesize The hyperfine splitting (leading order and
next to leading order) vs $\alpha_s(M_z)$ at $\mu=3$ GeV, the
unit is in MeV.}
\begin{ruledtabular}
\begin{tabular}{lccccc}
$\alpha_s(M_z)$&0.116&0.117&0.118&0.119&0.120 \\ \hline
$V_{hf}$&$49$&$54$&$58$&$63$&$69$\\
\hline $\delta V_{hf}$&$-7$&$-7$&$-8$&$-9$&$-10$
 \label{table3}
\end{tabular}
\end{ruledtabular}
\end{table*}

\begin{table*}[hbt]
\caption{\footnotesize The numerical results with different set of
$\mu$ at $\alpha_s(M_z)=0.118$, the unit is in MeV.}
\begin{ruledtabular}
\begin{tabular}{lccccc}
$\mu$&1&2&3&4&5 \\ \hline
$V_{hf}$&$101$&$69$&$58$&$52$&$49$\\\hline $\delta
V_{hf}$&$-81$&$-21$&$-8$&$-3$&$0.4$\\
 \hline $V_{hf}+\delta
V_{hf}$&$20$&$48$&$50$&$50$&$49$\\\hline
$V_{SI}$&$-$101&$-$75&$-$66&$-$61&$-$58\\\hline $ m_{\overline{\rm
MS}}$&$4.207$&$4.210$&$4.211$&$4.212$&$4.212$ \label{table4}
\end{tabular}
\end{ruledtabular}
\end{table*}

\acknowledgments
The work of C.S.K. was supported in part
by Grant No. R02-2003-000-10050-0 from BRP of the KOSEF.
The work of G.W. was supported in part
by Grant No. R02-2003-000-10050-0 from BRP of the KOSEF, and
in part by Grant No. F01-2004-000-10292-0 of KOSEF-NSFC International
Collaborative Research Grant. The work of T.L. is supported by BK21 program.

\end{document}